\documentclass{article}

\usepackage{PRIMEarxiv}

\usepackage[utf8]{inputenc} 
\usepackage[T1]{fontenc}    
\usepackage{hyperref}       
\usepackage{url}            
\usepackage{booktabs}       
\usepackage{amsfonts}       
\usepackage{nicefrac}       
\usepackage{microtype}      
\usepackage{lipsum}
\usepackage{fancyhdr}       
\usepackage{graphicx}       
\graphicspath{{media/}}     
\usepackage{amsmath} 
\usepackage{multirow} 
\usepackage[numbers]{natbib}
\bibpunct[, ]{(}{)}{;}{a}{}{,}
\renewcommand\bibfont{\fontsize{10}{12}\selectfont}

\usepackage{tabularx}
\usepackage{bm}

\title{A Bayesian Inference Approach for Reducing Inter-Investigator Variability in Sampling-Based Land Cover Classification
\thanks{\textit{The accepted manuscript is found at International Journal of Remote Sensing.}  
 Tsutsumida, N., \& Kato, A. (2025). A bayesian inference approach for reducing inter-investigator variability in sampling-based land cover classification. International Journal of Remote Sensing, 46(10), 4054–4077. https://doi.org/10.1080/01431161.2025.2496529}}

\author{
  Narumasa Tsutsumida \\
  Graduate school of Science \& Engineering,  \\
  Saitama University, Japan \\
  \texttt{rsnaru.jp@gmail.com} 
  \And
  Akira Kato \\
  Graduate School of Horticulture \\
  Chiba University, Japan 
}

\begin{document}
\maketitle

\begin{abstract}
Land cover classification faces persistent challenges with inter-investigator variability and salt-and-pepper noise. Although cloud platforms such as Google Earth Engine have made land cover classification more accessible, these issues persist, particularly when multiple investigators contribute to the process. This study developed a robust classification approach that integrates unsupervised clustering of investigator maps with a Bayesian inference framework using Dirichlet distributions. In this study, 44 investigators collected stratified reference samples across four land cover classes using point-based visual interpretation in Saitama City, Japan. We trained three different classifiers, Random Forests (RF), Support Vector Machines (SVM), and Single hidden layer Feed-forward Neural Networks (SFNN), and enhanced the system by implementing unsupervised clustering (k-Means or k-Medoids) to group reliable maps based on entropy characteristics. The Bayesian framework, employing Dirichlet distributions for both likelihood and prior distributions, enables sequential probability updates while preserving probabilistic class assignments. The Bayesian inference from the SVM classification maps achieved the highest mean overall accuracy of 0.857 for Monte Carlo sampling from the referenced JAXA land use land cover map, improving upon the non-Bayesian SVM map (0.855, p $<$ 0.001). Analysis revealed a strong correlation (r= 0.710) between investigators' labeling quality and classification accuracy, suggesting that selecting high-quality investigator maps improves the robustness of fusion. The Interspersion and Juxtaposition Index (IJI) showed that fused maps from SVM-based maps selected by k-Means reduced salt-and-pepper noise (IJI: 56.652) compared to baseline maps (IJI: 69.867). Our approach demonstrates an effective approach for combining multiple land cover classifications while reducing both inter-investigator variation and classification noise through integrated machine learning and Bayesian inference.

\end{abstract}


\section*{1 Introduction}
\label{Introduction}

Land cover classification plays a vital role in understanding and managing the Earth's terrestrial environment \citep{Wulder2018-pz, Pandey2021-ah}.
By categorising the Earth's surface into distinct, interpretable classes such as forests, urban areas, agricultural lands, and water bodies \citep{Ban2015-zi, Ramankutty2008-wz},
land cover maps serve as essential tools for various applications, including environmental monitoring, forest management, and urban planning \citep{Yang2021-ns, McGrath2015-is, Zeferino2020-li}.
Accurate and reliable land cover information is crucial not only for describing terrestrial surfaces but also for guiding sustainable development and informing policy decisions \citep{Barrow2006-aq}.

However, the process of land cover classification is inherently fraught with uncertainties \citep{Chen2022-up, Valle2023-yo, Zhang2009-yr}.
These uncertainties stem from various sources, such as data quality, classification algorithms, and temporal variations \citep{Foody2002-vr, Cheng2021-si}, and can lead to discrepancies in land cover maps, compromising their reliability and applicability.
A particularly overlooked aspect of this uncertainty is the variability introduced by investigators themselves. Inter-investigator variability in land cover classification represents a complex interplay of cognitive and methodological factors that significantly impacts classification outcomes \citep{Powell2004-qk, Hearn2011-ip, Morrison2016-jm, Cherrill1999-ye, Cherrill2016-ng, Cherrill1995-ao}. This variability manifests through two main mechanisms: interpretative discrepancies in land cover identification and differences in sampling decisions. The concept of inter-investigator variability has a rich history in scientific research, particularly in medical diagnostics \citep{Cochrane1952-vv, Yerushalmy1969-vp, Soloway1971-zq}. In land cover classification, this variability has unique characteristics because of the complexity of interpreting overhead imagery and landscape heterogeneity. Investigators often vary in their approaches for selecting training samples, including differences in quantity, spatial distribution, and representativeness.

Previous studies have demonstrated that inter-investigator variability leads to significant discrepancies in land cover maps. \citet{Cherrill1995-ao} found major differences between two field survey maps, with discrepancies covering 41.2\% of the surveyed area. These differences stem primarily from varying land cover interpretations among surveyors. \citet{Powell2004-qk} showed that trained interpreters disagreed on nearly 30\% of mixed reference samples, with most disagreements occurring in transitional classes. A questionnaire survey by \citet{Cherrill2016-ng} revealed that approximately 40\% of respondents had encountered erroneous reports. More recently, \citet{Reis2024-uf} documented substantial pixel-level disagreements between interpreters, even when analysing only five classes.
Nowadays, the use of very high-resolution satellite images for reference sampling has become increasingly common. \citet{Fritz2009-lp} developed a participatory geographic information system approach using Google Earth imagery to gather reference samples through a global investigator network. While these data helped validate and improve global land cover products \citep{See2015-wg}, sample quality varied significantly between experts and non-experts \citep{See2013-ah}. Further research has revealed the influence of cultural and national differences in the interpretation of landscape features \citep{Comber2015-tp}. These findings underscore the challenges of achieving consistent land cover interpretation through field surveys or visual sampling from high-resolution satellite imagery.

Despite the importance of addressing the impact of inter-investigator variability on land cover identification, with the emergence of user-friendly cloud geospatial platforms, such as Google Earth Engine (GEE), for referencing sampling as the ground truth, creating a land cover classification map has become easier.
GEE provides comprehensive tools for remote sensing analysis, including data loading and pre-processing, land cover classification, and mapping \citep{Amani2020-gk, Gorelick2017-fu}.
Land cover classification mapping using a Random Forest (RF) classifier or Support Vector Machine (SVM) on the GEE platform has been widely applied in research \citep{Tamiminia2020-zc, Zhao2021-gj, Perez-Cutillas2023-ln, Amani2020-gk} as well as for educational purposes \citep{Callejas2023-ea}.
Users can gather reference sample points using an interactive web map overlaid with high-resolution satellite images, even without following a formal sampling protocol.
This allows them to construct a classification model and perform mapping without downloading or processing data on their local computers.
Although land cover classification mapping has been easily achieved via such platforms, the importance of caution regarding inter-investigator variability is often overlooked.

Another challenge in uncertainty is the salt-and-pepper noise. Common machine learning algorithms such as RF and SVM, despite their popularity owing to their lightweight implementation, frequently produce noise—scattered isolated pixels that notably differ from their surroundings \citep{Hirayama2019-mo}. This noise degrades both the classification accuracy and the interpretability of visual maps in pixel-based classification tasks.
Multiple classifier systems (MCS) have emerged as an effective solution to reduce noise, thereby improving the accuracy by aggregating multiple classifiers \citep{Cruz2018-cb}. Some MCS approaches have been proposed in remote sensing perspective. For instance, \citet{Lopes2020-zs} reduced noise by developing a posterior fusion approach that combined different RF model results from time-series satellite imagery.
\citet{Mallinis2023-gk} applied some MCS approaches including plurality voting
algorithm to the classification results from different machine-learning models, resulting in a marginal increase in accuracy for binary classifications compared to individual classification models.
\citet{Hirayama2019-mo} demonstrated how an ensemble MCS approach could reduce salt-and-pepper noise by combining class probabilities from different supervised classifiers trained on the same dataset. 
\citet{Du2012-kg} applied a Bayesian average model that linearly combines the probabilistic predictions of multiple classifiers by weighting their posterior probabilities, achieving superior performance compared to other MCS approaches.
Some studies utilised a Latent Dirichlet Allocation (LDA) to fuse multiple categorical land cover maps \citep{Bahmanyar2018-wt, Li2021-sh}. The LDA is a generative probabilistic model that describes the latent structure of text document collections as distributions over discovered topics. As an unsupervised method, it primarily analyzes word co-occurrence patterns to identify thematic structures in textual data. Originally developed for text processing, LDA has been extended to fuse discrete thematic land cover classification maps by analysing co-occurrence patterns in categorical data. 
Thus, MCS is expected to reduce noise by combining different classification map results; however, these studies only investigated the applicability by using multiple classifications from a single training dataset.

While advances have been made in both understanding inter-investigator variability and developing methods to integrate multiple classification maps, these two challenges have largely been addressed separately. 
This gap is particularly overlooked, given that both challenges fundamentally relate to uncertainty in classification outcomes. While MCS approaches have demonstrated effectiveness in improving classification accuracy and reducing salt-and-pepper noise through the combination of multiple classifiers \citep{Mallinis2023-gk, Hirayama2019-mo}, their potential for reconciling inter-investigator variability has not been systematically investigated. 

Our study addresses this gap to create a consistent land cover classification map that deals with the uncertainties in sample-based mapping caused by inter-investigator variability. We propose a novel MCS approach that generates a unified classification map by fusing individual investigator maps to reduce uncertainty. We hypothesised that integrating multiple investigator maps through a Bayesian inference framework would mitigate individual biases and produce a more robust and accurate land cover map.
We made multiple classification maps by sampling data from different investigators using very high-resolution satellite images to simulate scenarios in which the data came from various web sources. Rather than enforcing a pre-defined sampling design, we maintained a flexible sampling protocol to demonstrate that our strategy can effectively integrate classification maps regardless of their source.

To test our approach, we focused on Saitama City, Japan, where we tasked 44 investigators with creating land cover maps using a consistent point-based visual interpretation scheme on GEE. We applied Bayesian inference with Dirichlet distribution to fuse the investigators maps. An unsupervised clustering was then employed to select accurate input maps for Bayesian inference to reduce inter-investigator variability in land cover mapping. This research highlights the importance of recognising and addressing intrinsic uncertainties in land cover classification, offering a promising solution to mitigate the impact of investigator variations.

\section*{2 Methods}

An overview of this approach is summarized in Figure 1.
Our approach to map fusion employs a Bayesian inference framework that combines multiple land cover maps generated by different investigators while accounting for inter-investigator variability. Three possible frameworks were considered:
(i) a basic Bayesian approach that fuses all investigator maps, 
(ii) a Bayesian approach that fuses all investigator maps with investigator-specific weights, and 
(iii) a basic Bayesian approach that fuses clustered investigator maps.
The third approach produces multiple outputs according to the number of clusters defined by unsupervised clustering (k-Means or k-Medoids). 

\begin{figure}[ht!]
\centering\includegraphics[width=1.0\linewidth]{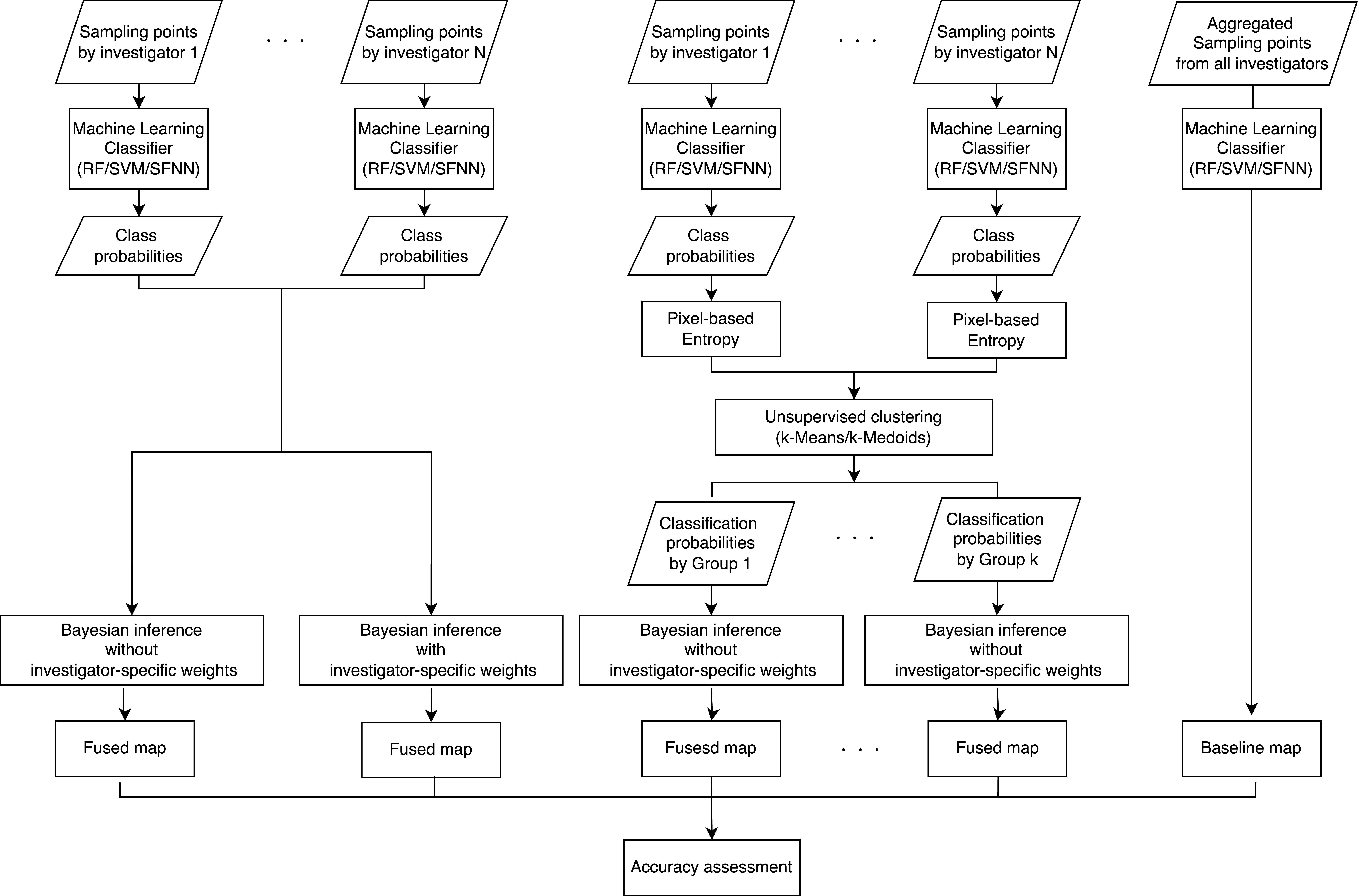}
\caption{ A flowchart of this study.}
\end{figure}

\subsection*{2.1 Study area}

We selected Saitama City, Japan, as our case study area.
With a population of 1.3 million people located in the greater Tokyo metropolitan area, Saitama City offers a blend of urban and suburban landscapes, featuring various residential, commercial, and green spaces, including agricultural areas spread throughout the region (Figure 2).

\begin{figure}[ht!]
\centering\includegraphics[width=1.0\linewidth]{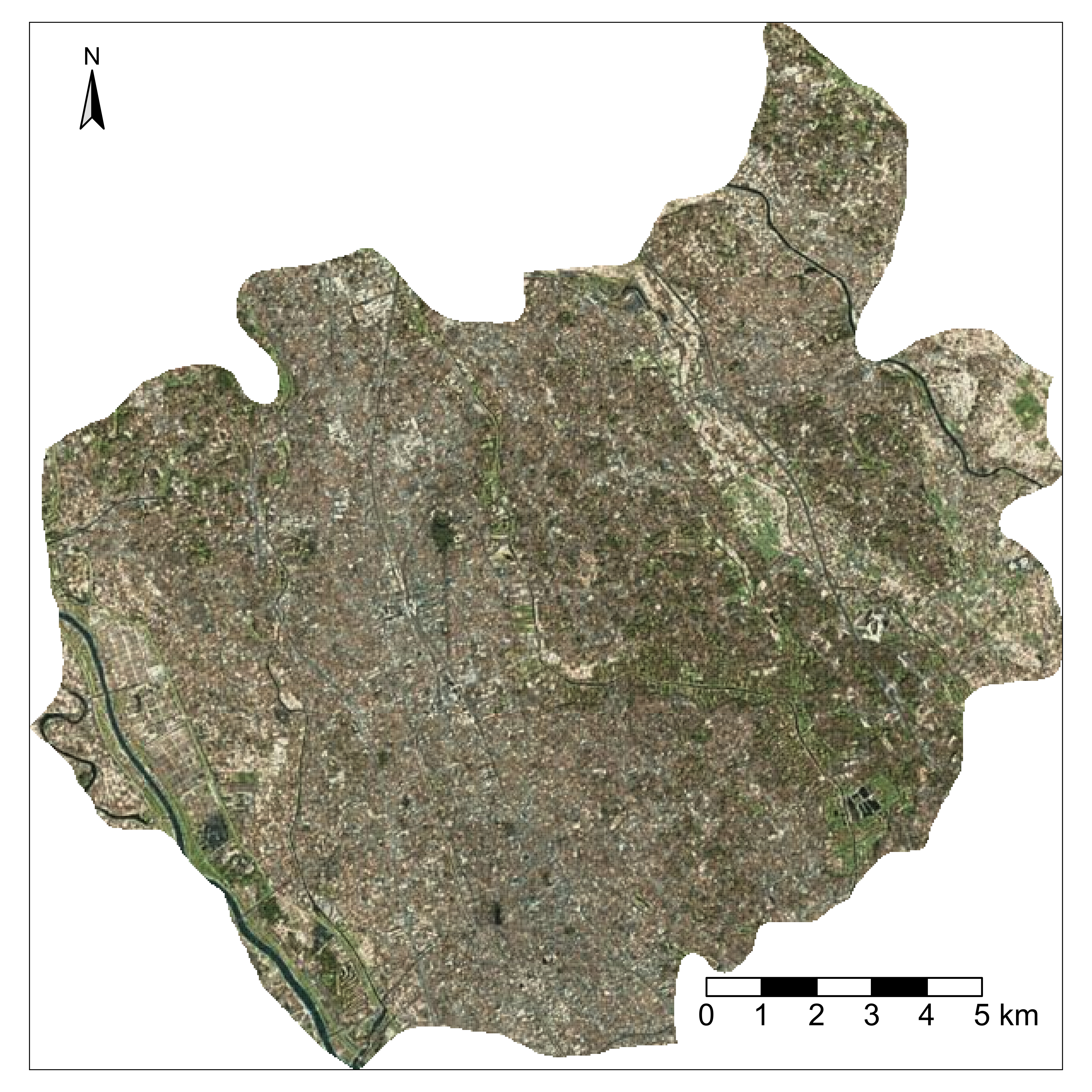}
\caption{Map of Saitama City, Japan as the study area. Data were obtained from the Bing map.}
\end{figure}

\subsection*{2.2 Data Collection and Preprocessing}

Forty-four independent investigators, undergraduate and postgraduate students, collected reference samples following instructions given in classes at Chiba University and Saitama University.
These classes provided an introduction to remote sensing and GEE.
The authors conducted hands-on lectures to teach investigators analytical techniques using the JavaScript playground in GEE.
Specifically, they used USGS Landsat 8 Level 2, Collection 2, Tier 1 data, which includes atmospherically corrected surface reflectance and land surface temperature derived from Landsat 8 OLI/TIRS sensors at a spatial resolution of 30 m.
The authors prepared an annual median composite of bands 1-7 and a normalised difference vegetation index (NDVI) for the study area in 2022.
Before creating the composite, clouds and snow cover were masked based on the QA\_PIXEL values.

Investigators collected reference data to create their own best accurate classification maps from the processed image to train a machine learning model. 
Before data collection began, we instructed investigators on the importance of high-quality reference data. They identified sample points across four distinct land cover strata classes: \textit{Forest}, \textit{Agriculture}, \textit{Urban}, and \textit{Water} using high-resolution Google satellite imagery on the GEE platform. 
While we established a minimum requirement of 10 sample points per class, investigators were strongly encouraged to collect additional samples to achieve their own best classification accuracy. 
The investigators collected a total of 8,011 training samples across the four land cover classes. On average, each investigator collected 42–48 samples per class, though this number varied between investigators (Figure 3).

\begin{figure}[ht!]
\centering\includegraphics[width=1.0\linewidth]{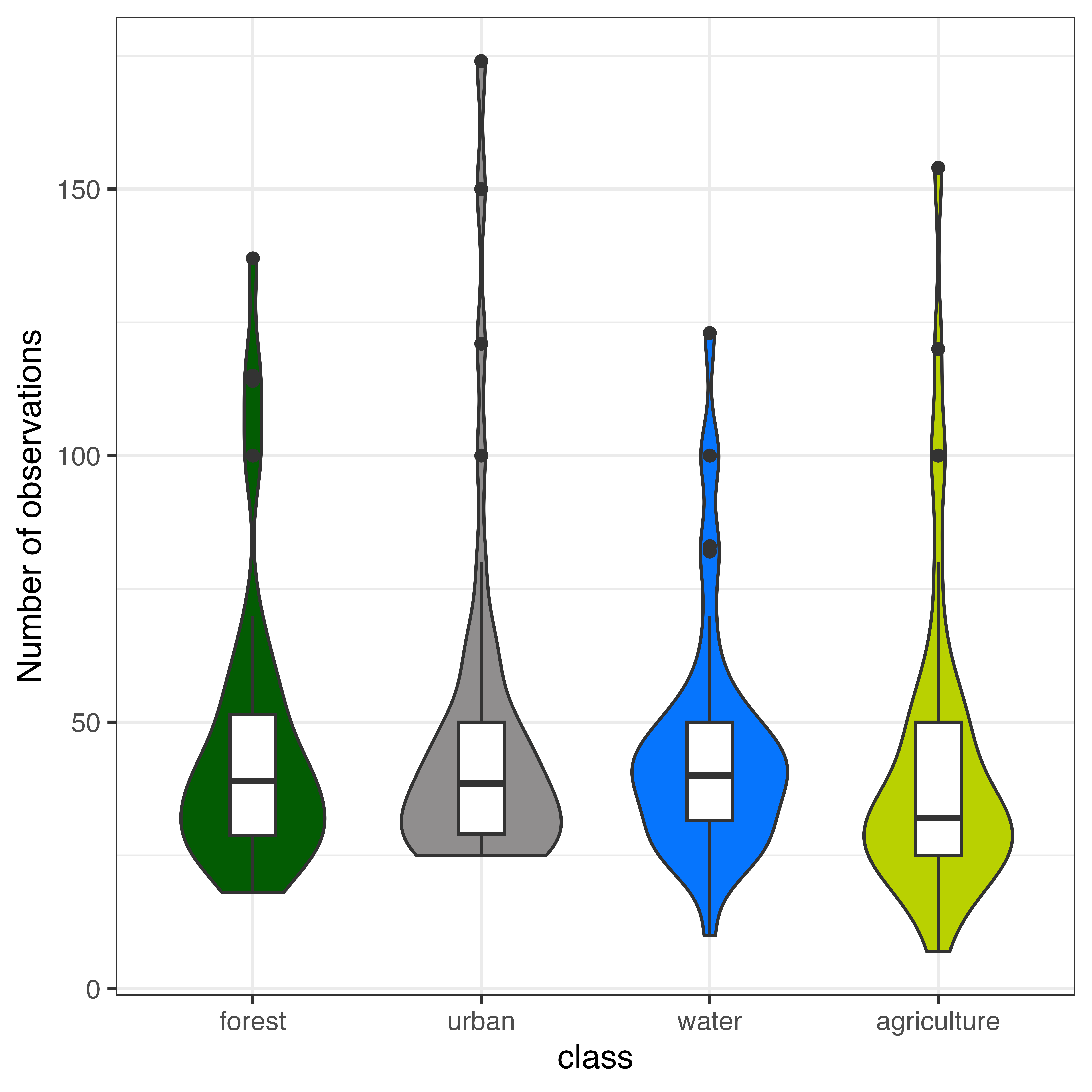}
\caption{Number of reference samples per class per investigator used in this study.}
\end{figure}

To investigate the inter-investigator variability in land cover classification and to develop a fusion approach that accounts for such variability, the research methodology incorporated decision tree, kernel, and neural network-based machine learning algorithms, each trained on investigator-provided samples to generate classification maps. To represent these three distinct algorithmic approaches, the experimental framework utilized RF, SVM, and Single-layer Feed-forward Neural Networks (SFNN), selected for their complementary architectural characteristics and proven effectiveness in remote sensing applications.
The RF implementation comprised a tree-based ensemble architecture optimized for non-linear relationship detection through recursive partitioning, configured with an ensemble of 1000 decision trees and 4 variables evaluated at each split node.
For the SVM implementation, we adopted a kernel-based approach that projected the feature space into higher dimensions to identify optimal decision boundaries. The algorithm employed a radial basis function kernel with hyperparameters set to cost=1 and gamma=0.1.
The SFNN architecture provided a neural network perspective through its capacity for hierarchical feature representation, implementing a hidden layer with 20 neurons. We deliberately chose this simpler architecture over deeper networks due to our limited training samples per investigator, which would likely cause overfitting. The three algorithms were independently trained on each investigator's dataset for land cover classification across the entire study region.

To establish a comparative baseline, we generated additional classification maps using the same three algorithms trained on the aggregated dataset without distinguishing between investigators' samples. 
These baseline classifications were analysed alongside the Bayesian inference results derived from the investigator-specific models.

\subsection*{2.3 Bayesian inference for map fusion}

We considered the above-mentioned three possible Bayesian inference frameworks.
The Dirichlet distribution was selected as the foundational probability distribution for our Bayesian inference framework because it models probability vectors over a finite set of categories, making it suitable for land cover classification where class probabilities must sum to one and remain non-negative.

\subsubsection*{2.3.1 Basic Bayesian fusion}

In the basic fusion approach, for each pixel location $i$, we model the observed class probabilities from each investigator using a Dirichlet distribution:

\begin{equation}
\mathbf{p}_{ij} \sim \text{Dirichlet}(\boldsymbol{\theta}_i)
\end{equation}

where $\mathbf{p}_{ij}$ represents the observed class probabilities from investigator $j$ at location $i$, and $\boldsymbol{\theta}_i$ is the vector of true class probabilities. A non-informative prior is placed on $\boldsymbol{\theta}_i$:

\begin{equation}
\boldsymbol{\theta}_i \sim \text{Dirichlet}(\mathbf{1})
\end{equation}

where $\mathbf{1}$ is a vector of ones, representing a uniform prior over the probability simplex. This formulation ensures maximum entropy without prior information while maintaining proper probability constraints.
Because the Dirichlet distribution requires strictly positive values, we substitute zero probabilities with $\epsilon = 1 \times 10^{-10}$ and normalize each probability set to sum to 1. 

\subsubsection*{2.3.2 Bayesian fusion with investigator-specific weights}

The weighted approach extends the basic model by introducing investigator-specific concentration parameters ($\kappa_j$) that serve as reliability weights. For each pixel location $i$ and investigator $j$, the model becomes:

\begin{equation}
\mathbf{p}_{ij} \sim \text{Dirichlet}(\kappa_j \boldsymbol{\theta}_i)
\end{equation}

where $\boldsymbol{\theta}_i$ is the true class probability vector at location $i$, $\kappa_j$ is the concentration parameter (weight) for investigator $j$,  $\mathbf{p}_{ij}$ is the vector of observed class probabilities from investigator $j$ at location $i$.

The priors are specified as:
\begin{align}
\boldsymbol{\theta}_i &\sim \text{Dirichlet}(\mathbf{1}) \\
\kappa_j &\sim \text{Gamma}(2, 1)
\end{align}

The $\kappa_j$  modulates the influence of each investigator's classifications. A higher $\kappa_j$ value indicates more substantial confidence in investigator $j$'s classifications. The Gamma(2,1) prior ensures positivity while providing sufficient flexibility for the data to determine appropriate weights.

\subsubsection*{2.3.3 Clustering of investigator maps}

The clustering stage employs k-Means and k-Medoids algorithms to partition investigators maps into coherent groups based on their characteristics. While k-Means minimizes the within-cluster sum of squared Euclidean distances, k-Medoids optimizes with respect to Manhattan distances and uses actual observations as cluster centres, providing robustness against outliers in the classification patterns.

Cluster assignments were determined using pixel-wise entropy as the feature space:

\begin{equation}
H(i) = -\sum_{c=1}^{n} P(x_c) \log_{2} P(x_c),
\end{equation}

where $H(l)$ represents the entropy of a land cover classification at location $i$, $P(x_c)$ represents the probability of the $c$-th class of land cover $x$, $n$ represents the total number of land cover classes (in this study, $n=4$), and $\log_{2}$ represents logarithm base 2. Entropy was selected as the feature space for clustering because it effectively represents the uncertainty in classification decisions as a single value for each pixel. Maps with similar entropy patterns reflect similar classification behavior and confidence levels among investigators. This makes entropy particularly suitable for grouping maps with coherent classification patterns using k-Means and k-Medoids algorithms prior to Bayesian fusion.

All groups in different $k$ ($k \in \{2,3,4\}$) clusters were tested as inputs for the Bayesian inference to evaluate their performance.

\subsubsection*{2.3.4 Final class determination}

For all fusion approaches (basic, weighted, and clustered), the final land cover class at each location is determined by selecting the class with the highest posterior probability:

\begin{equation}
\hat{y}_i = \arg\max_c E[\theta_{ic}|X_i]
\end{equation}

where $E[\theta_{ic}|X_i]$ is the posterior mean probability for class $c$ at location $i$.

The posterior distribution resulting from the Bayesian formulation follows a Dirichlet distribution due to the conjugate relationship between the Dirichlet prior and likelihood. Specifically, for each pixel location $i$, after observing the class probabilities from $J$ investigators, the posterior distribution is:

\begin{equation}
\theta_i | X_i \sim \text{Dirichlet}(\alpha + \sum_{j=1}^{J} w_j p_{ij})
\end{equation}

where $\alpha$ represents the prior parameters (set to $\mathbf{1}$ for non-informative prior), $p_{ij}$ represents the observed class probabilities from investigator $j$, and $w_j$ is the investigator-specific weight (equal to 1 in the unweighted approach, equal to $\kappa_j$ in the weighted approach). The expected value of this posterior, which provides the final fused probabilities, is calculated as:

\begin{equation}
E[\theta_{ic} | X_i] = \frac{\alpha_c + \sum_{j=1}^{J} w_j p_{ijc}}{\sum_{c'=1}^{C} \alpha_{c'} + \sum_{j=1}^{J} w_j}
\end{equation}

where $C$ is the total number of classes.

\subsection*{2.5 Assessments}

We assessed 44 investigator maps and fused maps using the Japan Aerospace Exploration Agency (JAXA) land use land cover map (JAXA-LULC map, ver. 23.12) \citep{Jaxa2018-ej}. The JAXA-LULC map is an accurate 14-class classification map with a spatial resolution of 10 m and an overall accuracy of 95.53\% across Japan. We resampled this to a 30 m resolution with our four target classes.

To ensure robust statistical validation, we employed Monte Carlo sampling with 100 iterations across the study area. In each iteration, we performed stratified random sampling to obtain 300 samples per land cover class from the JAXA-LULC map, resulting in 1,200 validation points per iteration. For each iteration, we computed overall accuracy, user's and producer's accuracies. 
The overall accuracy is the proportion of correct classifications across all classes relative to the total samples. 
The user's accuracy for class $c$ ($ua_{c}$) is the proportion of correct classifications for class $c$ relative to the total number of classifications for that class.
The producer's accuracy for class $c$ ($pa_{c}$) is the proportion of correct classifications for class $c$ relative to all the actual instances of that class, as shown in Equations (10)–(12):

\begin{equation}
\label{eq:oa}
oa = \frac{{\sum_{c}{n_{cc}}}}{{\sum_{c}{\sum_{q}{n_{cq}}}}},
\end{equation}

\begin{equation}
\label{eq:ua}
ua_{c} = \frac{{n_{cc}}}{{\sum_{q}{n_{cq}}}}
\end{equation}

\begin{equation}
\label{eq:pa}
pa_c = \frac{{n_{cc}}}{{\sum_{q}{n_{qc}}}}
\end{equation}

where the ${n_{cq}}$ refers to the number of instances classified as class $c$ that were actually from class $q$.

By using accuracy measures obtained by the Monte Carlo validation scheme, we conducted paired t-tests comparing each accuracy measure of each fused map against its corresponding baseline map to evaluate improvements in classification accuracy. 
Given the inherent limitations of achieving perfect spatial independence in evaluating land cover classification, particularly for classes with limited distribution (e.g., Water), we maintained equal sample sizes across classes to ensure fair comparison during statistical testing rather than compromising sample representativeness.

Furthermore, to evaluate the degree of salt-and-pepper noise, we applied a map-level Interspersion and Juxtaposition Index (IJI) to all the classification maps.
The IJI measures the intermixing of different patch types on a map. A higher IJI value indicates that patch types are more evenly interspersed, which corresponds to a more salt-and-pepper appearance, calculated by Equation (11):

\begin{equation}
\label{eq:iji}
I = -\frac{\sum_{a=1}^{m}\sum_{b=a+1}^{m} \left[ \left(\frac{e_{ab}}{E}\right) \ln \left(\frac{e_{ab}}{E}\right) \right]}{\ln\left(\frac{m(m-1)}{2}\right)} \times 100,
\end{equation}

where $e_{ab}$ is the total length of edge between patch types $a$ and $b$, $E$ is the total edge length in the landscape and $m$ is the number of patch types.

We selected these metrics for their interpretability as a land cover study. While the Kappa coefficient accounts for chance agreement, it has been criticized for its dependency on reference data prevalence \citep{Pontius2011-yi}. We opted for overall accuracy as our primary metric, particularly appropriate for our stratified random sampling of 300 points per class, ensuring unbiased evaluation across classes. User's and producer's accuracies were chosen over F1 scores to provide a more interpretable assessment of commission and omission errors for each land cover type. The IJI was specifically selected to quantify salt-and-pepper noise by measuring patch interspersion that is not captured by traditional accuracy metrics.

We conducted all analyses using R software version 4.1.2 \citep{R_Development_Core_Team2021-lh}. We performed the Bayesian analysis using the \textit{Rstan} package version 2.21.8 \citep{StanDevelopmentTeam2024-ui} and calculated the IJI using the \textit{landscapemetrics} package version 2.1.4 \citep{Hesselbarth2019-gd}.

\section*{3. Results}
\subsection*{3.1 Variability of individual investigators' maps}

Our analysis confirmed inconsistencies among classification maps created from individual investigators' samples. The Monte Carlo sampling analysis showed Pearson's correlation coefficients between investigator agreement ratios and mean overall accuracy of 0.672 for RF, 0.710 for SVM, and 0.781 for SFNN. These positive correlations across all three classifiers indicate that investigators whose reference samples had higher agreement with the JAXA reference data typically achieved better overall accuracy in their classification maps. This pattern persisted across different machine-learning models. This inter-investigator variability could affect the accuracy of land cover classification maps when combining them into a single unified map.

\begin{figure}[ht!]
\centering\includegraphics[width=1.0\linewidth]{Figs/Fig4_misinterpretation\_analysis.png}
\caption{ Scatter plots between agreement ratios in investigator labels against references and mean overall accuracies for prediction by (a) random forest, (b) support vector machine, and (c) single hidden layer feed-forward neural networks classifications in Monte Carlo sampling analyses over 100 iterations.}
\end{figure}

We found negative correlations between investigator agreement ratios and IJI values across the classification maps: -0.455 for RF, -0.274 for SVM, and -0.150 for SFNN. Higher sampling quality tends to produce less noisy classification maps, though SFNN shows limited ability to reduce salt-and-pepper errors regardless of sampling quality.

\begin{figure}[ht!]
\centering\includegraphics[width=1.0\linewidth]{Figs/Fig5_misinterpretation\_analysis_iji.png}
\caption{ Scatter plots between agreement ratios in investigator labels against references and map-level Interspersion and Juxtaposition Index (IJI) values of classification map by (a) random forest, (b) support vector machine, and (c) single hidden layer feed-forward neural networks models.}
\end{figure}

\subsection*{3.2 Map fusion}

To explore the impact of Bayesian fusion techniques from investigators' maps, we evaluated results by RF, SVM, and SFNN (Tables 1–3). Detailed summaries are plotted in Appendix Figures 1–3. Among the baseline models, SVM demonstrated the highest overall accuracy (0.855), followed by SFNN (0.838) and RF (0.836), with each classifier exhibiting distinct strengths in handling different land cover classes.

The fusion of all 44 maps produced statistically significant improvements in mean overall accuracy compared to the baseline across all three classifiers ($p <0.001$). This fusion approach resulted in reduced IJI values, demonstrating that the Bayesian fusion method effectively minimises classification noise while maintaining or improving classification accuracy. However, the weighted and unweighted model approaches showed similar performance in terms of accuracy measures and IJI values.

Applying clustering methods to the fusion process maintains or improves accuracy measures while reducing IJI values, with k-Medoids and k-Means clustering showing similar effectiveness. For RF, the second group of 4 k-Medoids categories (n=11) achieved the optimal configuration, with a mean overall accuracy of 0.851 and significant improvements across maps. This configuration showed particular strength in \textit{Water} classification (mean user's accuracy 0.961) and \textit{Urban} (mean producer's accuracy 0.794). The SVM classifier performed exceptionally well in several configurations: the first group of 2 k-Medoids categories (n=23), the second group of 2 k-Means categories (n=31), and the first group of 3 k-Means categories (n=11), all reaching mean overall accuracies of 0.857. These groups also showed notably reduced IJI values of 58.712, 61.074, and 56.652, respectively, compared to the baseline of 69.867. While the improvement in mean overall accuracy from 0.855 to 0.857 appears modest, the substantial reduction in IJI values from 69.867 in the baseline SVM model in the optimized k-Means clustered Bayesian fusion approach indicates a significant improvement in the spatial consistency, with fewer isolated pixels and more contiguous land cover patches that better represent actual landscape patterns. The SFNN reached its highest mean overall accuracy (0.854) in the second group of 3 k-Means categories (n=12), with a moderate reduction in IJI value from baseline (79.450 to 68.877). While the clustering approach with Bayesian inference improved both mean overall accuracy and IJI values across all classifiers, there was no clear advantage between k-Means and k-Medoids as the choice of unsupervised method.

\begin{table}[htbp]
\centering
\caption{Mean accuracy measures of random forest classifications by 100 Monte Carlo sampling and Interspersion Juxtaposition Index (IJI).}
\label{tab:stats_summary_rf}
\resizebox{\textwidth}{!}{
\begin{tabular}{llllllllllll}
\hline
\multirow{2}{*}{\textbf{Model}} & \textbf{n} & \textbf{Overall} & \multicolumn{4}{c}{\textbf{User's Accuracy}} & \multicolumn{4}{c}{\textbf{Producer's Accuracy}} & \textbf{IJI} \\
\cline{4-11}
& & \textbf{Accuracy} & \textbf{Forest} & \textbf{Agri.} & \textbf{Urban} & \textbf{Water} & \textbf{Forest} & \textbf{Agri.} & \textbf{Urban} & \textbf{Water} &  \\
\hline
Baseline & 44 & 0.836 & 0.852 & 0.745 & 0.903 & 0.849 & 0.874 & 0.711 & 0.777 & 0.983 & 78.688 \\
\hline
\multicolumn{12}{l}{\textit{All fused}} \\
Unweighted & 44 & 0.847*** & 0.875*** & 0.707*** & 0.882*** & 0.936*** & 0.903*** & 0.754*** & 0.746*** & 0.988*** & 64.374 \\
Weighted & 44 & 0.848*** & 0.882*** & 0.709*** & 0.878*** & 0.934*** & 0.897*** & 0.756*** & 0.752*** & 0.988*** & 64.199 \\
\hline
\multicolumn{12}{l}{\textit{k-Medoids}} \\
k2g1 & 25 & 0.845*** & 0.903*** & 0.680*** & 0.871*** & 0.958*** & 0.855*** & 0.796*** & 0.743*** & 0.985* & 56.691 \\
k2g2 & 19 & 0.830*** & 0.856** & 0.718*** & 0.889*** & 0.855*** & \textbf{0.928}*** & 0.671*** & 0.731*** & 0.990*** & 75.295 \\
k3g1 & 18 & 0.833** & 0.911*** & 0.648*** & 0.887*** & 0.955*** & 0.841*** & \textbf{0.828}*** & 0.683*** & 0.982 & 53.636 \\
k3g2 & 13 & 0.848*** & 0.866*** & 0.728*** & 0.833*** & 0.955*** & 0.914*** & 0.700*** & 0.792*** & 0.986*** & 62.520 \\
k3g3 & 13 & 0.828*** & 0.872*** & 0.736*** & 0.880*** & 0.823*** & 0.911*** & 0.656*** & 0.755*** & 0.992*** & 79.562 \\
k4g1 & 16 & 0.835 & \textbf{0.914}*** & 0.657*** & 0.871*** & 0.955*** & 0.824*** & 0.813*** & 0.724*** & 0.981** & 54.808 \\
k4g2 & 11 & \textbf{0.851}*** & 0.867*** & 0.731*** & 0.839*** & 0.961*** & 0.915*** & 0.711 & \textbf{0.794}*** & 0.985** & 60.144 \\
k4g3 & 11 & 0.826*** & 0.866*** & 0.732*** & 0.875*** & 0.827*** & 0.921*** & 0.645*** & 0.750*** & 0.989*** & 79.838 \\
k4g4 & 6 & 0.661*** & 0.704*** & 0.558*** & \textbf{0.906} & 0.615*** & 0.798*** & 0.641*** & 0.465*** & 0.740*** & 62.578 \\
\hline
\multicolumn{12}{l}{\textit{k-Means}} \\
k2g1 & 14 & 0.832*** & 0.884*** & 0.728*** & 0.888*** & 0.833*** & 0.904*** & 0.687*** & 0.748*** & 0.991*** & 76.995 \\
k2g2 & 30 & 0.846*** & 0.889*** & 0.688*** & 0.872*** & 0.958*** & 0.880*** & 0.782*** & 0.738*** & 0.984 & 57.878 \\
k3g1 & 5 & 0.809*** & 0.885*** & 0.613*** & 0.849*** & \textbf{0.969}*** & 0.827*** & 0.811*** & 0.641*** & 0.956*** & \textbf{48.157} \\
k3g2 & 26 & 0.850*** & 0.886*** & 0.702*** & 0.873*** & 0.953*** & 0.890*** & 0.770*** & 0.751*** & 0.987*** & 59.756 \\
k3g3 & 13 & 0.835 & 0.877*** & \textbf{0.740}** & 0.876*** & 0.842*** & 0.913*** & 0.671*** & 0.766*** & 0.990*** & 78.311 \\
k4g1 & 13 & 0.835 & 0.913*** & 0.654*** & 0.858*** & \textbf{0.969}*** & 0.819*** & 0.808*** & 0.736*** & 0.978*** & 51.435 \\
k4g2 & 8 & 0.794*** & 0.878*** & 0.692*** & 0.894*** & 0.745*** & 0.866*** & 0.636*** & 0.677*** & \textbf{0.996}*** & 76.055 \\
k4g3 & 17 & 0.846*** & 0.854 & 0.714*** & 0.879*** & 0.940*** & \textbf{0.928}*** & 0.728*** & 0.742*** & 0.986*** & 62.413 \\
k4g4 & 6 & 0.761*** & 0.701*** & 0.690*** & 0.867*** & 0.833*** & 0.893*** & 0.713 & 0.720*** & 0.717*** & 75.472 \\
\hline
\end{tabular}
}
\footnotesize{\textit{Note:} * $p<0.05$, ** $p<0.01$, *** $p<0.001$ compared to baseline model.}
\end{table}

\begin{table}[htbp]
\centering
\caption{Mean accuracy measures of support vector machine classifications by 100 Monte Carlo sampling and Interspersion Juxtaposition Index (IJI).}
\label{tab:stats_summary_svm}
\resizebox{\textwidth}{!}{%
\begin{tabular}{llllllllllll}
\hline
\multirow{2}{*}{\textbf{Model}} & \textbf{n} & \textbf{Overall} & \multicolumn{4}{c}{\textbf{User's Accuracy}} & \multicolumn{4}{c}{\textbf{Producer's Accuracy}} & \textbf{IJI} \\
\cline{4-11}
& & \textbf{Accuracy} & \textbf{Forest} & \textbf{Agri.} & \textbf{Urban} & \textbf{Water} & \textbf{Forest} & \textbf{Agri.} & \textbf{Urban} & \textbf{Water} &  \\
\hline
Baseline & 44 & 0.855 & 0.910 & 0.739 & \textbf{0.910} & 0.876 & 0.851 & 0.784 & 0.796 & 0.989 & 69.867 \\
\hline
\multicolumn{12}{l}{\textit{All fused}} \\
Unweighted & 44 & \textbf{0.857}*** & 0.879*** & 0.726*** & 0.885*** & 0.946*** & 0.901*** & 0.761*** & 0.778*** & 0.988 & 63.037 \\
Weighted & 44 & \textbf{0.857}*** & 0.881*** & 0.724*** & 0.884*** & 0.947*** & 0.898*** & 0.764*** & 0.778*** & 0.988 & 62.505 \\
\hline
\multicolumn{12}{l}{\textit{k-Medoids}} \\
k2g1 & 23 & \textbf{0.857}*** & 0.896*** & 0.710*** & 0.890*** & 0.954*** & 0.880*** & 0.794*** & 0.767*** & 0.988 & 58.712 \\
k2g2 & 21 & 0.850*** & 0.858*** & 0.753*** & 0.880*** & 0.902*** & 0.920*** & 0.701*** & 0.791** & 0.990*** & 72.697 \\
k3g1 & 12 & 0.856 & 0.914*** & 0.699*** & 0.878*** & \textbf{0.962}*** & 0.845*** & 0.808*** & 0.783*** & 0.988* & 54.569 \\
k3g2 & 16 & 0.848*** & 0.881*** & 0.721*** & 0.895*** & 0.902*** & 0.901*** & 0.745*** & 0.756*** & 0.990* & 70.816 \\
k3g3 & 16 & 0.853* & 0.841*** & 0.755*** & 0.880*** & 0.927*** & \textbf{0.938}*** & 0.696*** & 0.790*** & 0.988* & 68.680 \\
k4g1 & 11 & 0.855 & \textbf{0.917}*** & 0.694*** & 0.880*** & \textbf{0.962}*** & 0.841*** & \textbf{0.816}*** & 0.778*** & 0.987*** & \textbf{53.834} \\
k4g2 & 10 & 0.830*** & 0.899*** & 0.718*** & 0.899*** & 0.819*** & 0.888*** & 0.707*** & 0.733*** & 0.991*** & 74.923 \\
k4g3 & 9 & 0.854 & 0.849*** & 0.730*** & 0.882*** & 0.960*** & 0.915*** & 0.743*** & 0.774*** & 0.985*** & 64.585 \\
k4g4 & 14 & 0.846*** & 0.860*** & 0.756*** & 0.878*** & 0.882*** & 0.911*** & 0.691*** & 0.792* & 0.990*** & 72.304 \\
\hline
\multicolumn{12}{l}{\textit{k-Means}} \\
k2g1 & 13 & 0.842*** & 0.863*** & 0.759*** & 0.879*** & 0.858*** & 0.901*** & 0.683*** & 0.792* & 0.990** & 74.101 \\
k2g2 & 31 & \textbf{0.857}** & 0.887*** & 0.717*** & 0.886*** & 0.950*** & 0.890*** & 0.776*** & 0.773*** & 0.988 & 61.074 \\
k3g1 & 11 & \textbf{0.857}*** & 0.904*** & 0.701*** & 0.895*** & 0.959*** & 0.866*** & 0.813*** & 0.762*** & 0.988 & 56.652 \\
k3g2 & 13 & 0.842*** & 0.863*** & 0.759*** & 0.879*** & 0.858*** & 0.901*** & 0.683*** & 0.792* & 0.990** & 74.101 \\
k3g3 & 20 & 0.855 & 0.876*** & 0.728*** & 0.878*** & 0.941*** & 0.904*** & 0.749*** & 0.780*** & 0.988* & 64.580 \\
k4g1 & 8 & 0.819*** & 0.865*** & \textbf{0.778}*** & 0.873*** & 0.771*** & 0.866*** & 0.615*** & \textbf{0.802}*** & \textbf{0.992}*** & 81.358 \\
k4g2 & 11 & 0.854 & 0.858*** & 0.734** & 0.885*** & 0.939*** & 0.914*** & 0.736*** & 0.778*** & 0.988* & 67.453 \\
k4g3 & 10 & 0.854 & 0.908* & 0.696*** & 0.898*** & 0.949*** & 0.855*** & 0.813*** & 0.756*** & 0.988 & 58.657 \\
k4g4 & 15 & 0.856 & 0.875*** & 0.725*** & 0.873*** & 0.954*** & 0.904*** & 0.753*** & 0.780*** & 0.987*** & 60.445 \\
\hline
\end{tabular}
}

\footnotesize{\textit{Note:} * $p<0.05$, ** $p<0.01$, *** $p<0.001$ compared to baseline model.}
\end{table}

\begin{table}[htbp]
\centering
\caption{Mean accuracy measures of single hidden layer feed-forward neural networks classifications by 100 Monte Carlo sampling and Interspersion Juxtaposition Index (IJI).}
\label{tab:stats_summary_nnet}
\resizebox{\textwidth}{!}{%
\begin{tabular}{llllllllllll}
\hline
\multirow{2}{*}{\textbf{Model}} & \textbf{n} & \textbf{Overall} & \multicolumn{4}{c}{\textbf{User's Accuracy}} & \multicolumn{4}{c}{\textbf{Producer's Accuracy}} & \textbf{IJI} \\
\cline{4-11}
& & \textbf{Accuracy} & \textbf{Forest} & \textbf{Agri.} & \textbf{Urban} & \textbf{Water} & \textbf{Forest} & \textbf{Agri.} & \textbf{Urban} & \textbf{Water} &  \\
\hline
Baseline & 44 & 0.838 & 0.845 & 0.724 & 0.900 & 0.894 & 0.890 & 0.755 & 0.731 & 0.974 & 79.450 \\
\hline
\multicolumn{12}{l}{\textit{All fused}} \\
Unweighted & 44 & 0.852*** & 0.844 & 0.742*** & 0.896*** & 0.932*** & 0.906*** & 0.749*** & 0.778*** & 0.976* & 73.891 \\
Weighted & 44 & 0.851*** & 0.844 & 0.742*** & 0.894*** & 0.927*** & 0.906*** & 0.748*** & 0.777*** & 0.973 & 74.220 \\
\hline
\multicolumn{12}{l}{\textit{k-Medoids}} \\
k2g1 & 34 & 0.853*** & 0.860*** & 0.730** & 0.887*** & 0.941*** & 0.886* & 0.767*** & 0.785*** & 0.972*** & 69.851 \\
k2g2 & 10 & 0.835** & 0.805*** & 0.747*** & 0.914*** & 0.883*** & 0.930*** & 0.696*** & 0.730 & 0.984*** & 83.658 \\
k3g1 & 23 & 0.850*** & 0.873*** & 0.714*** & 0.880*** & 0.949*** & 0.867*** & \textbf{0.783}*** & 0.782*** & 0.968*** & \textbf{65.689} \\
k3g2 & 11 & 0.839 & 0.832*** & 0.758*** & 0.901 & 0.865*** & 0.910*** & 0.698*** & 0.765*** & 0.982*** & 85.040 \\
k3g3 & 10 & 0.827*** & 0.767*** & 0.749*** & 0.904** & 0.900*** & \textbf{0.949}*** & 0.645*** & 0.733 & 0.980*** & 82.689 \\
k4g1 & 22 & 0.851*** & 0.873*** & 0.716*** & 0.879*** & \textbf{0.951}*** & 0.867*** & 0.780*** & 0.788*** & 0.969*** & 65.727 \\
k4g2 & 11 & 0.842*** & 0.794*** & 0.759*** & 0.905*** & 0.917*** & 0.941*** & 0.686*** & 0.759*** & 0.982*** & 82.075 \\
k4g3 & 10 & 0.823*** & 0.765*** & \textbf{0.762}*** & 0.900 & 0.878*** & 0.940*** & 0.633*** & 0.735 & 0.984*** & 86.910 \\
k4g4 & 1 & 0.784*** & \textbf{0.881}*** & 0.700*** & 0.815*** & 0.756*** & 0.826*** & 0.695*** & 0.719*** & 0.897*** & 78.665 \\
\hline
\multicolumn{12}{l}{\textit{k-Means}} \\
k2g1 & 19 & 0.832*** & 0.777*** & 0.752*** & 0.905** & 0.909*** & 0.943*** & 0.679*** & 0.731 & 0.976* & 83.293 \\
k2g2 & 25 & 0.853*** & 0.854*** & 0.735*** & 0.893*** & 0.933*** & 0.893 & 0.758 & 0.783*** & 0.976*** & 71.806 \\
k3g1 & 19 & 0.827*** & 0.764*** & 0.748*** & 0.866*** & 0.938*** & 0.916*** & 0.685*** & 0.755*** & 0.951*** & 79.217 \\
k3g2 & 12 & \textbf{0.854}*** & 0.854*** & 0.733*** & 0.884*** & \textbf{0.951}*** & 0.892 & 0.763*** & 0.790*** & 0.971*** & 68.877 \\
k3g3 & 13 & 0.827*** & 0.845 & 0.738*** & \textbf{0.919}*** & 0.822*** & 0.897*** & 0.692*** & 0.729 & \textbf{0.990}*** & 84.960 \\
k4g1 & 10 & 0.852*** & 0.877*** & 0.717*** & 0.888*** & 0.942*** & 0.856*** & 0.780*** & \textbf{0.791}*** & 0.982*** & 68.685 \\
k4g2 & 11 & 0.813*** & 0.738*** & 0.749*** & 0.845*** & 0.927*** & 0.900*** & 0.638*** & 0.761*** & 0.952*** & 84.767 \\
k4g3 & 15 & 0.830*** & 0.832*** & 0.726 & 0.917*** & 0.858*** & 0.909*** & 0.711*** & 0.714*** & 0.985*** & 80.992 \\
k4g4 & 8 & 0.846*** & 0.826*** & 0.752*** & 0.875*** & 0.934*** & 0.924*** & 0.724*** & 0.781*** & 0.957*** & 74.712 \\
\hline
\end{tabular}
}
\footnotesize{\textit{Note:} * $p<0.05$, ** $p<0.01$, *** $p<0.001$ compared to baseline model.}
\end{table}

To complement the quantitative analysis, we conducted a visual assessment of the classification maps (Figures 6 and 7). Figure 6 presents a comparative analysis of maps generated by RF, SVM, and SFNN, including baseline models, Bayesian inference results from all 44 investigators' maps, and maps produced from k-Medoids clustering with optimal accuracy and minimal IJI values. The spatial distribution of land cover classes showed consistent patterns across all three classification models, indicating that the choice of classifier influences the final output characteristics independently of the Bayesian inference application.

The baseline models for RF, SVM, and SFNN (see details in Figure 7(\textit{b}, \textit{f}, \textit{j})) showed significant classification noise. In contrast, the Bayesian fusion maps created from all investigator maps achieved both improved accuracy and reduced noise. Maps generated from clustered inputs with the highest mean overall accuracy (see details in Figure 7(\textit{d}, \textit{h}, \textit{l}) demonstrated better alignment with the reference map, especially in boundary definition and class distribution. The maps optimized for minimum IJI values (see details in Figure 7\textit{e}, \textit{i}, \textit{m})) showed the most uniform distribution of land cover classes, though this spatial uniformity slightly decreased classification accuracy.

\begin{figure}[ht!]
\centering\includegraphics[width=1.0\linewidth]{Figs/Fig6_Whole_maps_arrange.png}
\caption{  Examples of classification results: (\textit{a}) Referenced land cover map by Japan Aerospace Exploration Agency (JAXA); land cover classification maps by (\textit{b}) Random Forest (RF) from all samples, (\textit{c}) Bayesian inference from all RF classification maps, (\textit{d}) Bayesian inference with highest overall accuracy from the second group of 4 k-Medoids clustered ($k4-2$) RF maps, (\textit{e}) Bayesian inference with lowest Interspersion Juxtaposition Index (IJI)  from the first group of 3 k-Means clustered ($k3-1$) RF maps, (\textit{f}) Support Vector Machine (SVM) map from all samples, (\textit{g}) Bayesian inference from all SVM maps, (\textit{h}) Bayesian inference with highest overall accuracy from the first group of 2 k-Medoids clustered ($k2-1$) SVM maps,
(\textit{i}) Bayesian inference with lowest IJI from the first group of 4 k-Medoids clustered ($k4-1$) SVM maps, (\textit{j}) Single hidden layer Feed-forward Neural Networks (SFNN) map from all samples, (\textit{k}) Bayesian inference from all SFNN maps, (\textit{l}) Bayesian inference with highest overall accuracy from the second group of 3 k-Means clustered ($k3-2$) SFNN maps, and (\textit{m}) Bayesian inference with lowest IJI from the first group of 4 k-Medoids clustered ($k4-1$) SFNN maps. The red rectangles indicate the area shown in Figure 8. }
    \end{figure}

\begin{figure}[ht!]
\centering\includegraphics[width=1.0\linewidth]{Figs/Fig7_SampleArea_maps_arrange_r1.png}
\caption{ Detailed maps of the sample site marked by a red rectangle in Figure 6: (\textit{a}) Referenced land cover map by Japan Aerospace Exploration Agency (JAXA); land cover classification maps by (\textit{b}) Random Forest (RF) from all samples, (\textit{c}) Bayesian inference from all RF classification maps, (\textit{d}) Bayesian inference with highest overall accuracy from the second group of 4 k-Medoids clustered ($k4-2$) RF maps, (\textit{e}) Bayesian inference with lowest Interspersion Juxtaposition Index (IJI)  from the first group of 3 k-Means clustered ($k3-1$) RF maps, (\textit{f}) Support Vector Machine (SVM) map from all samples, (\textit{g}) Bayesian inference from all SVM maps, (\textit{h}) Bayesian inference with highest overall accuracy from the first group of 2 k-Medoids clustered ($k2-1$) SVM maps,
(\textit{i}) Bayesian inference with lowest IJI from the first group of 4 k-Medoids clustered ($k4-1$) SVM maps, (\textit{j}) Single hidden layer Feed-forward Neural Networks (SFNN) map from all samples, (\textit{k}) Bayesian inference from all SFNN maps, (\textit{l}) Bayesian inference with highest overall accuracy from the second group of 3 k-Means clustered ($k3-2$) SFNN maps, and (\textit{m}) Bayesian inference with lowest IJI from the first group of 4 k-Medoids clustered ($k4-1$) SFNN maps. }

\end{figure}

\section*{4 Discussion}

\subsection*{4.1  Impact of inter-investigator variability}

Inter-investigator variability is a critical issue in the construction of accurate land cover classification maps. This variability arises because of different subjective interpretations, methodological approaches, and levels of expertise among investigators, leading to inconsistencies in identifying and classifying land cover classes. In this study, we focused only on the variability of the collection process of reference samples by investigators using GEE. The quality and reliability of classification maps can differ substantially based on the investigator's choice of locations, number of samplings, and interpretations of land cover, depending on their knowledge, experience, and personal biases. We found large variability in the sampling quality among different investigators, and such quality tends to have a positive relation to classification accuracy (Figure 4). This indicates the importance of sampling quality for making a classification map from multiple investigators, but inter-investigator variability is inevitable, leading to considerable uncertainty in classification maps.

\subsection*{4.2 Reducing the impact of inter-investigator variability}

To address inter-investigator variability, we examined the Bayesian inference framework. The fused map produced by this framework improved accuracy and reduced salt-and-pepper noise by integrating class probabilities from different maps. Among three approaches tested, clustering proved most effective for improving accuracy and reducing noise. This suggests that Bayesian inference works best when fusing maps with similar characteristics, which the clustering approach helps identify by grouping similar maps from the candidate pool. The Bayesian inference approach using investigator-specific weights showed no clear improvement in fusion techniques, likely due to high inter-investigator variability in the input maps, failing to deal with such variability in the model. Our results showed that classification method choice significantly influenced the fused output—SVM performed best among the classifiers, while SFNN performed worst in terms of overall accuracy. While our clustering-based fusion approach applied to SVM showed marginal but statistically significant ($p < 0.001$) improvement in terms of overall accuracy, it produces more spatially consistent land cover patterns that better represent real-world land cover distributions, as visually confirmed in Figures 6 and 7. These findings highlight the importance of carefully selecting classifiers before fusing investigators' maps.

\subsection*{4.3 Limitations of this study}

First, this study focused on investigator-defined stratified sampling for pixel-based classification in Saitama City, Japan. While we demonstrated our approach for fusing land cover classification maps from different sources, we haven't tested its application to other sampling methods like random sampling. The broader applicability of our findings needs validation across different geographical contexts, since our approach may perform differently across varied landscapes and regional characteristics that would affect both interpretation and sampling variability among investigators. Regions with higher landscape heterogeneity might amplify inter-investigator variability, while cultural differences in landscape interpretation \citep{Comber2015-tp} could affect clustering effectiveness. Future research should test this framework across diverse biomes, settlement patterns, and cultural contexts to establish its global applicability.

Second, a restricted number of land cover classes was defined in this study. The classification was limited to only four categories: \textit{Forest}, \textit{Agriculture}, \textit{Urban}, and \textit{Water}. Although this simplification facilitates the sampling scheme and reduces confusion among investigators, it poses challenges in accurately representing the diversity and complexity of terrestrial environments when applied to more detailed classification schemes with a larger number of classes. Clearly, it is impossible to comprehensively delineate actual land cover into these four broad categories. For instance, bare lands and grasslands may be inappropriately allocated to the \textit{Agriculture} class. Moreover, many areas are composed of a mosaic of different land cover types within a 30-meter grid cell, with complex transitional zones and gradients of land use. Urban areas often contain substantial green spaces, such as parks and gardens, while agricultural lands can include patches of forests or other natural habitats. The performance of our Bayesian framework might differ when applied to more complex taxonomies with numerous sub-classes. Future studies should explore the scalability and performance of this approach when applied to more complex classification schemes.

Finally, our approach remains technically exploratory. The use of unsupervised clustering in this study assumes that the number of clusters reflects the dissimilarities in land cover classifications among investigators. The grouping of investigators' maps with similar characteristics offers a significant advantage; however, the sensitivity to the number of clusters (k) requires careful consideration. Increasing k tends to raise the similarity of map characteristics within a group, as it allows for a more fine-grained separation of investigators' maps based on their unique features. However, this comes at the cost of reducing the number of maps in each group, which may affect the robustness of the Bayesian inference applied to each cluster. The selection of the optimal k value likely depends on landscape characteristics, classification complexity, and investigator expertise levels, suggesting context-specific adaptation may be necessary. Future studies should seek a better optimized approach to find such hyperparameters toward automated processing that can adapt to different application contexts and investigator pools.

\section*{5 Conclusions}

As land cover classification mapping becomes increasingly accessible through cloud platforms like Google Earth Engine, addressing inter-investigator variability has become a critical concern. This study developed a novel framework that simultaneously reduces both inter-investigator variability and salt-and-pepper noise in land cover classification mapping. The Bayesian inference framework, which incorporates Dirichlet distributions for both likelihood and prior distributions, achieved accurate classification results with reduced noise. The unsupervised clustering approach for selecting input maps further outperformed the non-clustering Bayesian approach in both accuracy and noise reduction. This improvement remained consistent across all tested classifiers. Our approach provides a practical solution for addressing inter-investigator variability, particularly when implementing large-scale mapping projects involving multiple investigators. Future work will examine diverse case studies to validate the framework's robustness.

\section*{Acknowledgements}
We gratefully acknowledge all investigators who contributed to this work. We extend our appreciation to the two reviewers whose insightful feedback significantly enhanced the quality of this manuscript.

\section*{Funding}
This work was supported by Japan Science Technology Agency (ACT-X No.21455106) and JSPS KAKENHI (20K20005).

\section*{Data Availability Statement}
The data supporting the findings of this study are available from the corresponding author upon request.

\bibliographystyle{plainnat}
\bibliography{biblio}

\section*{Appendix}

\begin{figure}[ht!]
\centering\includegraphics[width=1.0\linewidth]{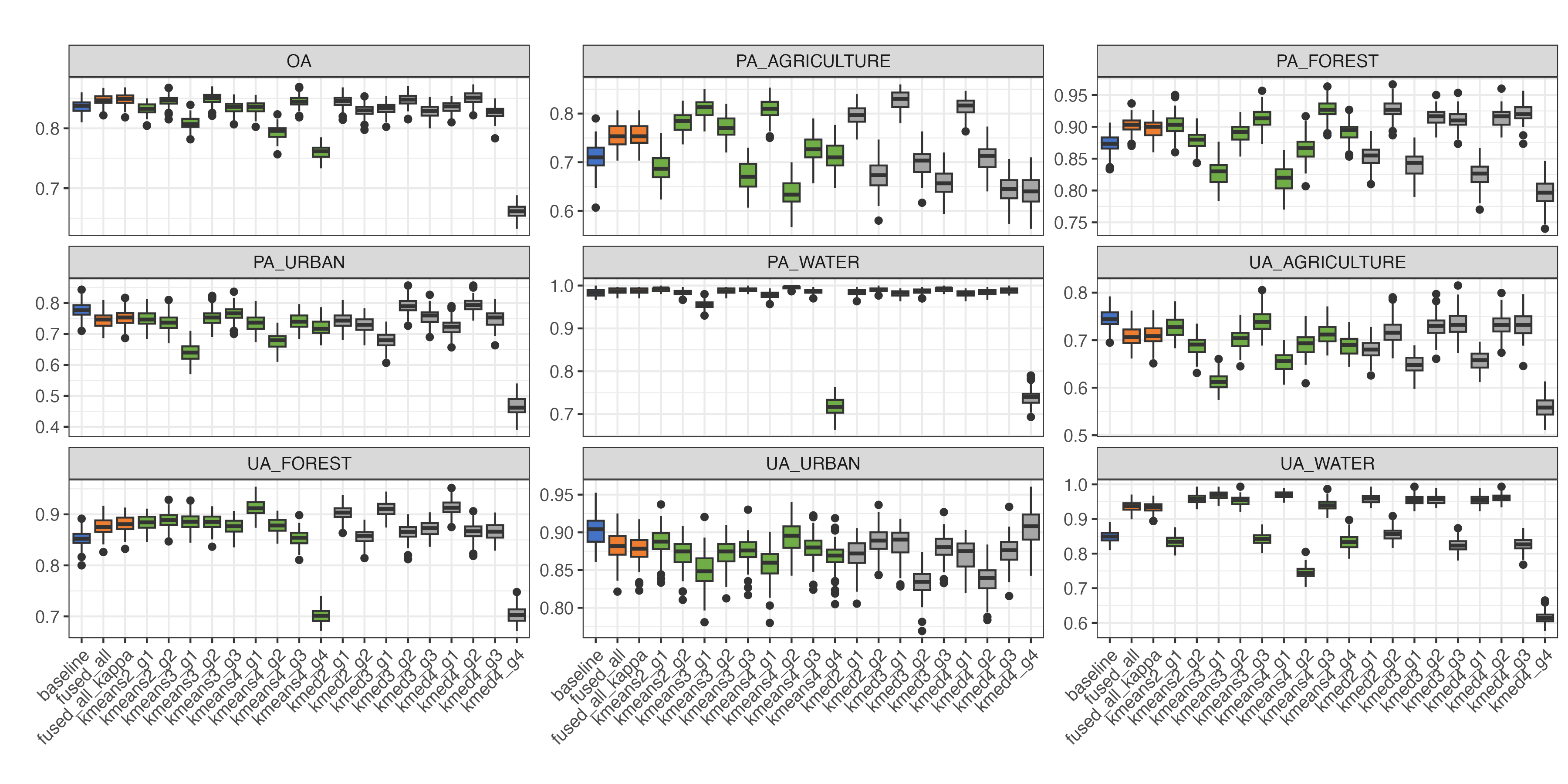}
\caption{ Boxplots of accuracy measures of random forest classifications by 100 Monte Carlo sampling.}
\end{figure}

\begin{figure}[ht!]
\centering\includegraphics[width=1.0\linewidth]{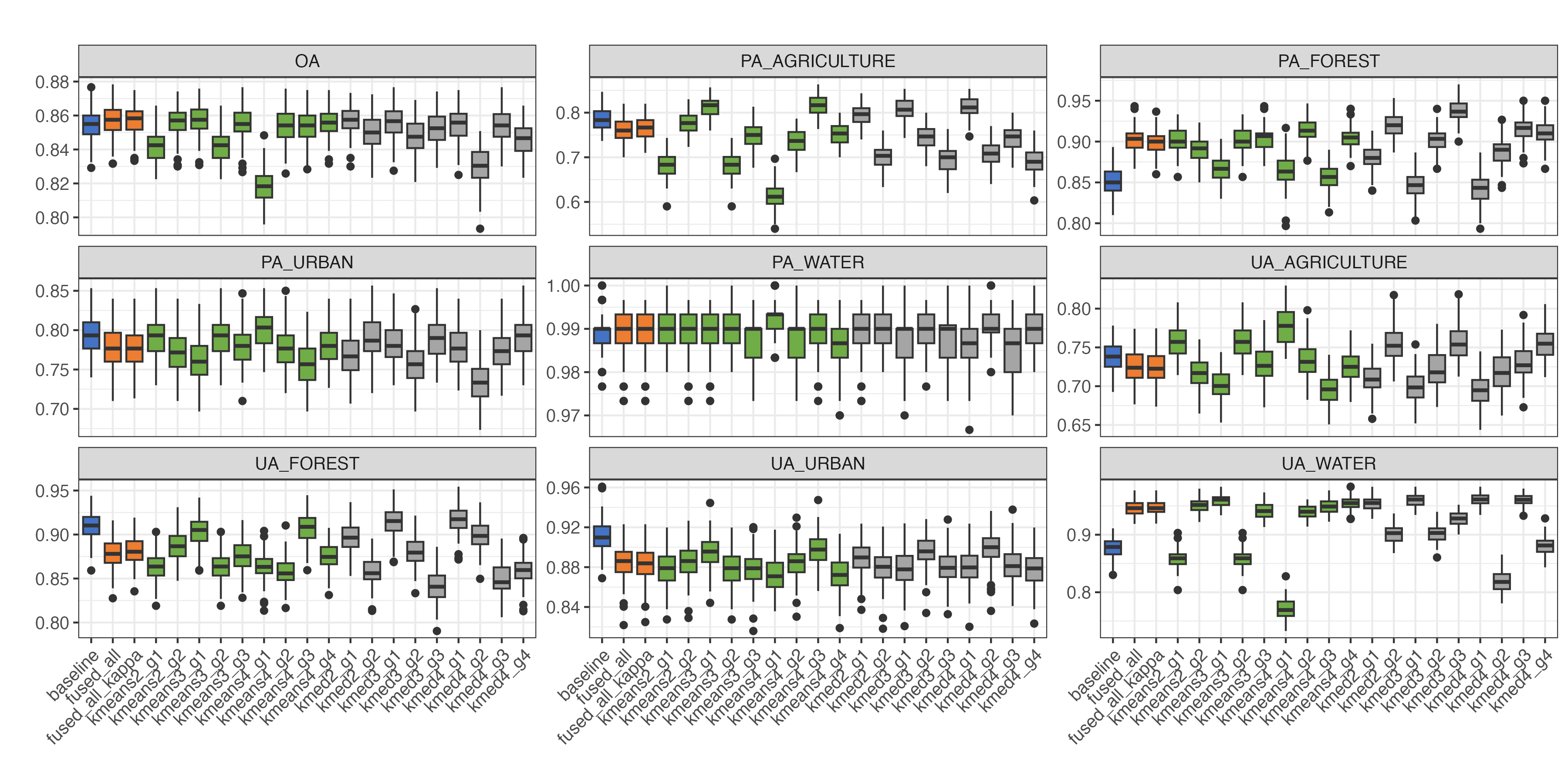}
\caption{ Boxplots of accuracy measures of support vector machine classifications by 100 Monte Carlo sampling.}
\end{figure}

\begin{figure}[ht!]
\centering\includegraphics[width=1.0\linewidth]{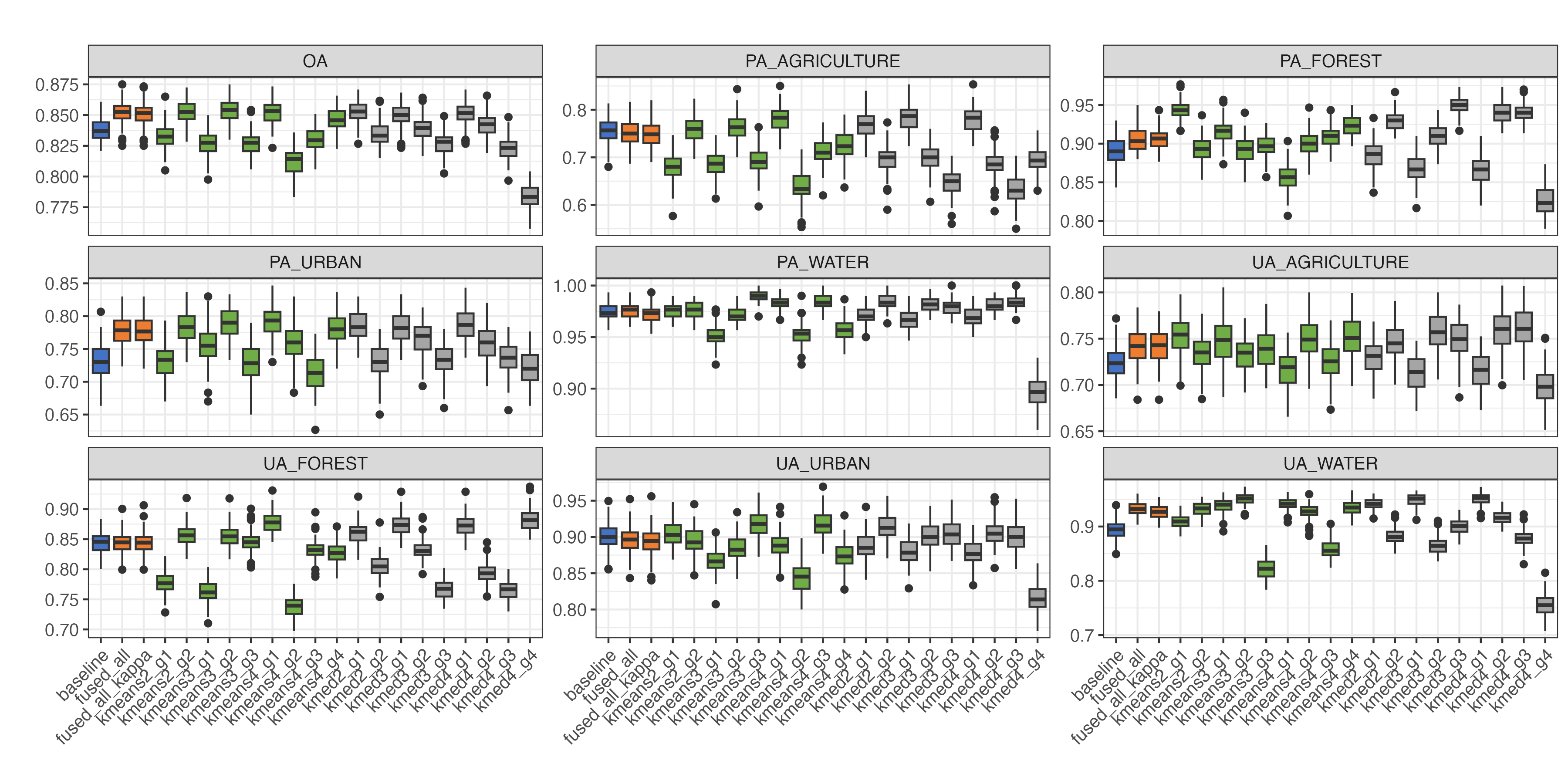}
\caption{ Boxplots of accuracy measures of single hidden layer feed-forward neural networks classifications by 100 Monte Carlo sampling.}
\end{figure}

\end{document}